\documentclass[twocolumn,aps,pra,amsmath,amssymb]{revtex4-1}
\usepackage[latin9]{inputenc}
\usepackage{amstext}
\usepackage{amssymb}
\usepackage{graphicx}
\usepackage{esint}

\makeatletter
\usepackage{amssymb}
\usepackage{epsfig}
\usepackage{bm}
\usepackage{dcolumn}
\usepackage{color}

\makeatother

\begin{document}

\title{Quantum and thermal fluctuations in a Raman spin-orbit coupled Bose
gas}

\author{Xiao-Long Chen}

\affiliation{Centre for Quantum and Optical Science, Swinburne University of Technology,
Melbourne, Victoria 3122, Australia}

\author{Xia-Ji Liu}

\affiliation{Centre for Quantum and Optical Science, Swinburne University of Technology,
Melbourne, Victoria 3122, Australia}

\author{Hui Hu}

\affiliation{Centre for Quantum and Optical Science, Swinburne University of Technology,
Melbourne, Victoria 3122, Australia}

\date{\today}
\begin{abstract}
We theoretically study a three-dimensional weakly-interacting Bose
gas with Raman-induced spin-orbit coupling at finite temperature.
By employing a generalized Hartree-Fock-Bogoliubov theory with Popov
approximation, we determine a complete finite-temperature phase diagram
of three exotic condensation phases (i.e., the stripe, plane-wave
and zero-momentum phases), against both quantum and thermal fluctuations.
We find that the plane-wave phase is significantly broadened by thermal
fluctuations. The phonon mode and sound velocity at the transition
from the plane-wave phase to the zero-momentum phase are thoughtfully
analyzed. At zero temperature, we find that quantum fluctuations open
an unexpected gap in sound velocity at the phase transition, in stark
contrast to the previous theoretical prediction of a vanishing sound
velocity. At finite temperature, thermal fluctuations continue to
significantly enlarge the gap, and simultaneously shift the critical
minimum. For a Bose gas of $^{87}$Rb atoms at the typical experimental
temperature, $T=0.3T_{0}$, where $T_{0}$ is the critical temperature
of an ideal Bose gas without spin-orbit coupling, our results of gap
opening and critical minimum shifting in the sound velocity, are qualitatively
consistent with the recent experimental observation {[}S.-C. Ji \textit{et
al.}, Phys. Rev. Lett. \textbf{114}, 105301 (2015){]}.
\end{abstract}


\maketitle
The interaction of a particle's spin with its spatial motion ­­­­-
the so-called spin-orbit coupling (SOC) - plays an important role
in various areas of physics. Over the past few years, the SOC effects
have been under extensive studies in alkali atomic quantum gases \cite{Dalibard2011,Galitski2013,Goldman2014,Zhai2015},
owing to the high controllability of cold-atom platforms \cite{Bloch2008}.
By utilizing two counter-propagating Raman lasers via a two-photon
process, physicists have realized one-dimensional SOC with an equal
weight combination of Rashba and Dresselhaus types, which couples
two atomic internal states \cite{Lin2011,Wang2012,Cheuk2012}. Most
recently, two-dimensional SOC of Rashba type is also achieved \cite{Huang2016,Wu2016}.
These seminal breakthroughs lead to fruitful researches both theoretically
\cite{Wang2010,Wu2011,Ho2011,Hu2012,Li2012PRL,Li2012EPL,Ozawa2012PRA,Martone2012,Zheng2012,Ozawa2012PRL,Zheng2013,Yu2013,Chen2017}
and experimentally \cite{Zhang2012,Ji2014,Ji2015,Li2017}, giving
rise to many fascinating phenomena, such as the exotic bosonic superfluidity
\cite{Wang2010,Wu2011,Ho2011,Hu2012,Li2012PRL,Yu2013}, topological
superfluidity and Majorana fermions \cite{Sato2009,Jiang2011,Liu2012a,Liu2012b,Liu2013,Liu2014},
and the spin Hall effect \cite{Zhu2006,Liu2007,Beeler2013}.

\begin{figure}
\centering{}\includegraphics[width=0.48\textwidth]{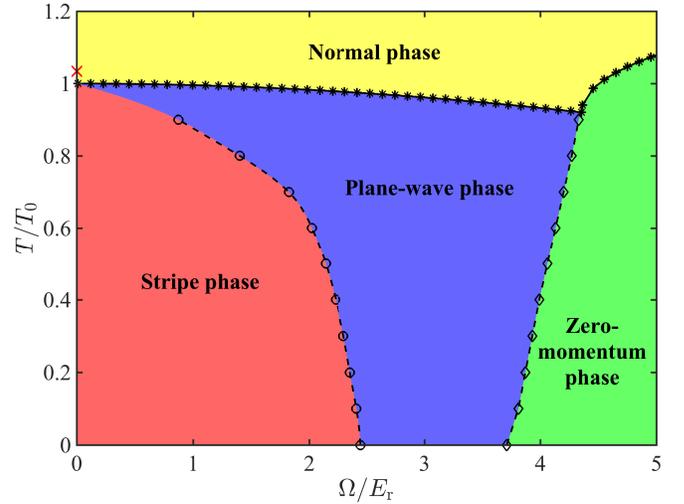}\caption{\label{fig1_soc} (color online). Phase diagram of a two-component
Bose gas with Raman induced spin-orbit coupling in the plane of temperature
$T$ and Raman Rabi frequency $\Omega$. The temperature is measured
in units of the critical BEC temperature of an ideal spinless Bose
gas with density $n/2$, i.e., $T_{0}=2\pi\hbar^{2}[(n/2)/\zeta(3/2)]^{2/3}/(mk_{B})$.
We take the total density of our SOC Bose gas, $n=1.0k_{\textrm{r}}^{3}$,
the intra-species interaction energy, $gn=0.8E_{\textrm{r}}$, and
the inter-species interaction energy, $g_{_{\uparrow\downarrow}}n=0.5E_{\textrm{r}}$.
Here, $k_{\textrm{r}}$ and $E_{\textrm{r}}=\hbar^{2}k_{\textrm{r}}^{2}/(2m)$
are Raman wave-vector and the recoil energy, respectively.}
\end{figure}

In this Rapid Communication, we are interested in a three-dimensional
weakly interacting Bose gas with a one-dimensional SOC at \emph{finite}
temperature. The zero-temperature phase diagram of such a Bose gas
was explored in detail in last few years \cite{Lin2011,Ho2011,Li2012PRL,Ozawa2012PRA,Martone2012,Zheng2013},
by using the Gross-Pitaevskii (GP) equation and the Bogoliubov theory.
There are typically three exotic phases by tuning the Rabi frequency
of the two Raman beams (see Fig. 1 at $T=0$): the stripe (ST) phase,
the plane-wave (PW) phase and the zero-momentum (ZM) phase \cite{Lin2011,Ho2011}.
Only a handful of theoretical investigations have addressed the finite-temperature
effects so far \cite{Ozawa2012PRL,Yu2014}, which are however unavoidable
in realistic experiments \cite{Ji2014,Ji2015}. Theoretically, Ozawa
and Baym discussed the stability of condensates against quantum and
thermal fluctuations \cite{Ozawa2012PRL}. Experimentally, by measuring
the magnetization of the condensate of $^{87}$Rb atoms, Ji \textit{et
al.} determined a qualitative finite-temperature phase diagram of
the ST-PW transition \cite{Ji2014}. A follow-up theoretical study
by Yu perturbatively calculated the ST-PW boundary at finite temperature
in terms of small Rabi frequency, and obtained a good agreement \cite{Yu2014}.
Unfortunately, this perturbation approach fails at relatively high
temperature, and also is not applicable for a large difference in
intra- and inter-species interaction strengths, which leads to a large
critical Rabi frequency \cite{Li2012PRL,Li2017}. Furthermore, the
phase transition from the PW phase to the ZM phase in the presence
of quantum and thermal fluctuations has not been explored both theoretically
and experimentally. The purpose of our work is to present a solid
calculation of a SOC Bose gas at finite temperature, by developing
a self-consistent Hartree-Fock-Bogoliubov theory within Popov approximation.
In this way, we address the quantum and thermal effects on different
condensation phases.

Our main results are briefly summarized in Fig. \ref{fig1_soc}, where
we delineate a finite-temperature phase diagram on Raman Rabi frequency
(the horizontal axis) and nonzero temperature (the vertical axis).
Four distinct regimes could be clearly identified: the ST phase (red),
PW phase (blue), ZM phase (green) and normal phase (yellow). We find
that, by increasing temperature, the PW phase is significantly enlarged
before reaching the superfluid-to-normal phase transition. In addition
to confirming the previous perturbative prediction by Yu \cite{Yu2014},
our results clearly reveal that the PW-ZM boundary is significantly
shifted by nonzero temperature, in sharp contrast to the naive picture
drawn earlier \cite{Ji2014}. This shift is also found in the sound
velocity, which exhibits a minimum at the PW-ZM transition (see Fig.
4). Therefore, it can be experimentally measured by using Bragg spectroscopy.
Indeed, for $^{87}$Rb atoms, our calculation with a realistic temperature
indicates a sizable shift in the sound velocity minimum, consistent
with the recent experimental measurement \cite{Ji2015} (see the inset
of Fig. 4).

\textit{Model}. A two-component Bose gas with Raman induced spin-orbit
coupling can be well described by the model Hamiltonian, $\hat{H}=\hat{H}_{0}+\hat{H}_{\textrm{int}}$
\cite{Li2012PRL,Martone2012,Zheng2013}, where the single-particle
Hamiltonian is ($\hbar=1$) 
\begin{equation}
\hat{H}_{0}=\int\mathrm{d}^{3}\mathbf{r}\left[\hat{\Phi}_{\uparrow}^{\dagger}(\mathbf{r}),\hat{\Phi}_{\downarrow}^{\dagger}(\mathbf{r})\right]\mathcal{H}_{s}(\hat{{\bf p}})\left[\begin{array}{c}
\hat{\Phi}_{\uparrow}(\mathbf{r})\\
\hat{\Phi}_{\downarrow}(\mathbf{r})
\end{array}\right],\label{eq:Hsp}
\end{equation}
with $\mathcal{H}_{s}(\hat{{\bf p}})=(\hat{{\bf p}}-k_{\textrm{r}}\mathbf{e}_{x}\sigma_{z})^{2}/(2m)+(\Omega\sigma_{x}+\delta\sigma_{z})/2$
and $\hat{{\bf p}}=-i\mathbf{\nabla}$, and the interaction Hamiltonian
is 
\begin{equation}
\hat{H}_{\textrm{int}}=\frac{1}{2}\int\mathrm{d}^{3}\mathbf{r}\sum_{\sigma,\sigma^{\prime}=\uparrow\downarrow}g_{\sigma\sigma^{\prime}}\hat{\Phi}_{\sigma}^{\dagger}\hat{\Phi}_{\sigma^{\prime}}^{\dagger}\hat{\Phi}_{\sigma^{\prime}}\hat{\Phi}_{\sigma}(\mathbf{r}),\label{eq:Hint}
\end{equation}
with $g_{\sigma\sigma^{\prime}}=4\pi a_{\sigma\sigma'}/m$. Here,
$\Omega$ and $\delta$ are respectively the Rabi frequency and detuning
of the Raman lasers, $\sigma_{x}$ and $\sigma_{z}$ are Pauli matrices,
and $a_{\sigma\sigma'}$ are the intra- ($\sigma=\sigma'$) and inter-species
($\sigma\neq\sigma'$) $s$-wave scattering lengths. The recoil momentum
$\mathbf{k}=k_{\textrm{r}}\mathbf{e}_{x}$ of the laser beams is along
the $x$-axis and the corresponding recoil energy is $E_{\textrm{r}}=k_{\textrm{r}}^{2}/(2m)$.
For $^{87}$Rb atoms, the SOC term $\hat{p}_{x}\sigma_{z}k_{\textrm{r}}/m$
in $\mathcal{H}_{s}(\hat{{\bf p}})$ describes a momentum-dependent
coupling between hyperfine states in the $F=1$ ground state manifold
\cite{Lin2011}. Following the typical experimental parameters \cite{Ji2014,Ji2015},
we assume a zero laser detuning, $\delta=0$, and the interaction
strengths $g_{\uparrow\uparrow}=g_{\downarrow\downarrow}=g\neq g_{\uparrow\downarrow}$.

\textit{Hartree-Fock-Bogoliubov-Popov theory}. To describe quantum
and thermal fluctuations, we generalize a Hartree-Fock-Bogoliubov
theory \cite{Griffin1996,Dodd1998,Buljan2005} within Popov approximation
\cite{PopovBook} (HFBP) and apply it to the SOC Bose gas. Following
the standard procedure \cite{Chen2015}, we separate the Bose field
operator $\hat{\Phi}({\bf r},t)$ into a combination of condensate
wave-functions $\phi_{\sigma}(\mathbf{r})$ and the fluctuation operators
$\hat{\eta_{\sigma}}(\mathbf{r},t)$ as ($\sigma=\uparrow,\downarrow$),
\begin{equation}
\left[\begin{array}{c}
\hat{\Phi}_{\uparrow}\\
\hat{\Phi}_{\downarrow}
\end{array}\right]=e^{-i\mu t}e^{iP_{x}x}\left[\sqrt{n_{c}}\left(\begin{array}{c}
\cos\theta\\
-\sin\theta
\end{array}\right)+\left(\begin{array}{c}
\hat{\eta}_{\uparrow}\\
\hat{\eta}_{\downarrow}
\end{array}\right)\right],\label{eq:plane_wave_ansatz}
\end{equation}
with the chemical potential $\mu$ and a uniform density $n_{c}=N_{c}/V$
\cite{Li2012PRL,Martone2012,Zheng2013}. For simplicity, in this work
we focus on a \emph{plane-wave} \textit{\emph{wave-function}} for
the condensate moving along the $x$-direction with the momentum $P_{x}\geq0$.
We also use an angle $\theta$ in the range $[0,\pi/4]$ to weight
the spin-up and spin-down condensate components. Both variational
parameters, $\theta$ and $P_{x}$, are to be determined by minimizing
the free energy of the system. The fluctuation operators $\hat{\eta}_{\sigma}$
and their conjugate can be expanded in a quasiparticle basis ($\hat{a}^{\dagger}$,
$\hat{a}$) under a Bogoliubov transformation, i.e., $\hat{\eta}_{\sigma}(\mathbf{r},t)=\sum_{j}[u_{j\sigma}({\bf r})e^{-i\omega_{j}t}\hat{a}_{j}+v_{j\sigma}^{*}({\bf r})e^{i\omega_{j}t}\hat{a}_{j}^{\dagger}]$,
where in the uniform case the quasiparticle amplitudes may take the
form, $u_{j\sigma}({\bf r})=u_{{\bf q}\sigma}^{(\tau)}\mathrm{exp}(i{\bf q\cdot r})$
and $v_{j\sigma}({\bf r})=v_{{\bf q}\sigma}^{(\tau)}\mathrm{exp}(i{\bf q\cdot r})$,
and the index $j=(\mathbf{q},\tau=\pm)$ of the quasiparticle energy
levels can be represented by the momentum $\mathbf{q}$ and the helicity
branch index $\tau=\pm$ due to the SOC \cite{Cao2014}.

By substituting the Bose field operators Eq. (\ref{eq:plane_wave_ansatz})
into the equations of motion $i\partial_{t}\hat{\Phi}_{\sigma}({\bf r},t)=[\hat{\Phi}_{\sigma},\hat{H}]$,
taking the standard mean-field decoupling (i.e., Hartree-Fock-Bogoliubov
approximation) \cite{Griffin1996}, and omitting the terms with anomalous
densities $n_{a}=\langle\hat{\eta}_{\sigma}\hat{\eta}_{-\sigma}\rangle$
(i.e., Popov approximation) \cite{PopovBook,Chen2015}, we obtain
two separate equations for the condensate and Bogoliubov quasiparticles,
respectively: (i) the modified GP equations, 
\begin{equation}
\left[\mathcal{H}_{s}(\hat{{\bf p}})+\mathrm{diag}(\mathcal{L}_{\uparrow},\mathcal{L}_{\downarrow})\right]\phi=\mu\phi,\label{eq:gp}
\end{equation}
with $\phi=(\phi_{\uparrow},\phi_{\downarrow})^{T}$ and $\mathcal{L}_{\sigma}\equiv g(n_{c\sigma}+2n_{t\sigma})+g_{_{\uparrow\downarrow}}n_{-\sigma}$,
and (ii) the generalized Bogoliubov equations,
\begin{eqnarray}
\left[\mathcal{H}_{s}\left(\hat{{\bf p}}+{\bf q}\right)-\mu+\mathcal{A}_{\uparrow}\right]U_{\mathbf{q}}^{(\tau)}+\mathcal{B}V_{\mathbf{q}}^{(\tau)} & = & \varepsilon_{\mathbf{q\tau}}U_{\mathbf{q}}^{(\tau)},\label{eq:bogo1}\\
-\mathcal{B}U_{\mathbf{q}}^{(\tau)}-\left[\mathcal{H}_{s}\left(\hat{{\bf p}}-{\bf q}\right)-\mu+\mathcal{A}_{\downarrow}\right]V_{\mathbf{q}}^{(\tau)} & = & \varepsilon_{\mathbf{q}\tau}V_{\mathbf{q}}^{(\tau)},\label{eq:bogo2}
\end{eqnarray}
where $U_{{\bf q}}^{(\tau)}=[u_{{\bf q}\uparrow}^{(\tau)},u_{{\bf q}\downarrow}^{(\tau)}]^{T}$,
$V_{{\bf q}}^{(\tau)}=[v_{{\bf q}\uparrow}^{(\tau)},v_{{\bf q}\downarrow}^{(\tau)}]^{T}$,
and
\begin{eqnarray}
\mathcal{A}_{\sigma} & \equiv & \left[\begin{array}{cc}
2gn_{\sigma}+g_{_{\uparrow\downarrow}}n_{-\sigma} & g_{_{\uparrow\downarrow}}\phi_{\sigma}\phi_{-\sigma}\\
g_{_{\uparrow\downarrow}}\phi_{\sigma}\phi_{-\sigma} & 2gn_{-\sigma}+g_{_{\uparrow\downarrow}}n_{\sigma}
\end{array}\right],\label{eq: VarAs}\\
\mathcal{B} & \equiv & \left[\begin{array}{cc}
g\phi_{\uparrow}^{2} & g_{_{\uparrow\downarrow}}\phi_{\uparrow}\phi_{\downarrow}\\
g_{_{\uparrow\downarrow}}\phi_{\uparrow}\phi_{\downarrow} & g\phi_{\downarrow}^{2}
\end{array}\right].\label{eq: VarB}
\end{eqnarray}
Here, $n_{c\uparrow}=n_{c}\cos^{2}\theta$ and $n_{c\downarrow}=n_{c}\sin^{2}\theta$
are the condensate density of the two components, $n_{t\sigma}\equiv\langle\hat{\eta}_{\sigma}^{\dagger}\hat{\eta}_{\sigma}\rangle=(1/V)\sum_{{\bf q\tau}}[(|u_{{\bf q}\sigma}^{(\tau)}|^{2}+|v_{{\bf q}\sigma}^{(\tau)}|^{2})/(e^{\beta\varepsilon_{\mathbf{q\tau}}}-1)+|v_{{\bf q}\sigma}^{(\tau)}|^{2}]$
with $\beta\equiv1/(k_{B}T)$ and $n_{\sigma}=n_{c\sigma}+n_{t\sigma}$
are the non-condensate density and the total density for spin $\sigma$,
respectively. We note that, the generalized Bogoliubov equations give
rise to four solutions with energies $\pm\varepsilon_{\mathbf{q},\tau=\pm}$.
We retain the two physical solutions with positive energies $\varepsilon_{\mathbf{q},\tau=\pm}>0$
only. Note also that, in all the previous studies, the non-condensate
density $n_{t\sigma}$ is neglected \cite{Li2012PRL,Martone2012,Zheng2013}.
This treatment is reasonable at zero temperature, where the \emph{quantum}
depletion $n_{t\sigma}^{(\textrm{Q})}=(1/V)\sum_{{\bf q\tau}}|v_{{\bf q}\sigma}^{(\tau)}|^{2}\sim0.01n_{_{\sigma}}$
is typically small in the weak-coupling regime. However, as we shall
see, near the phase transition the small quantum depletion may lead
to qualitative changes to some experimental observables such as sound
velocity.

\begin{figure}[t]
\centering{}\includegraphics[width=0.48\textwidth]{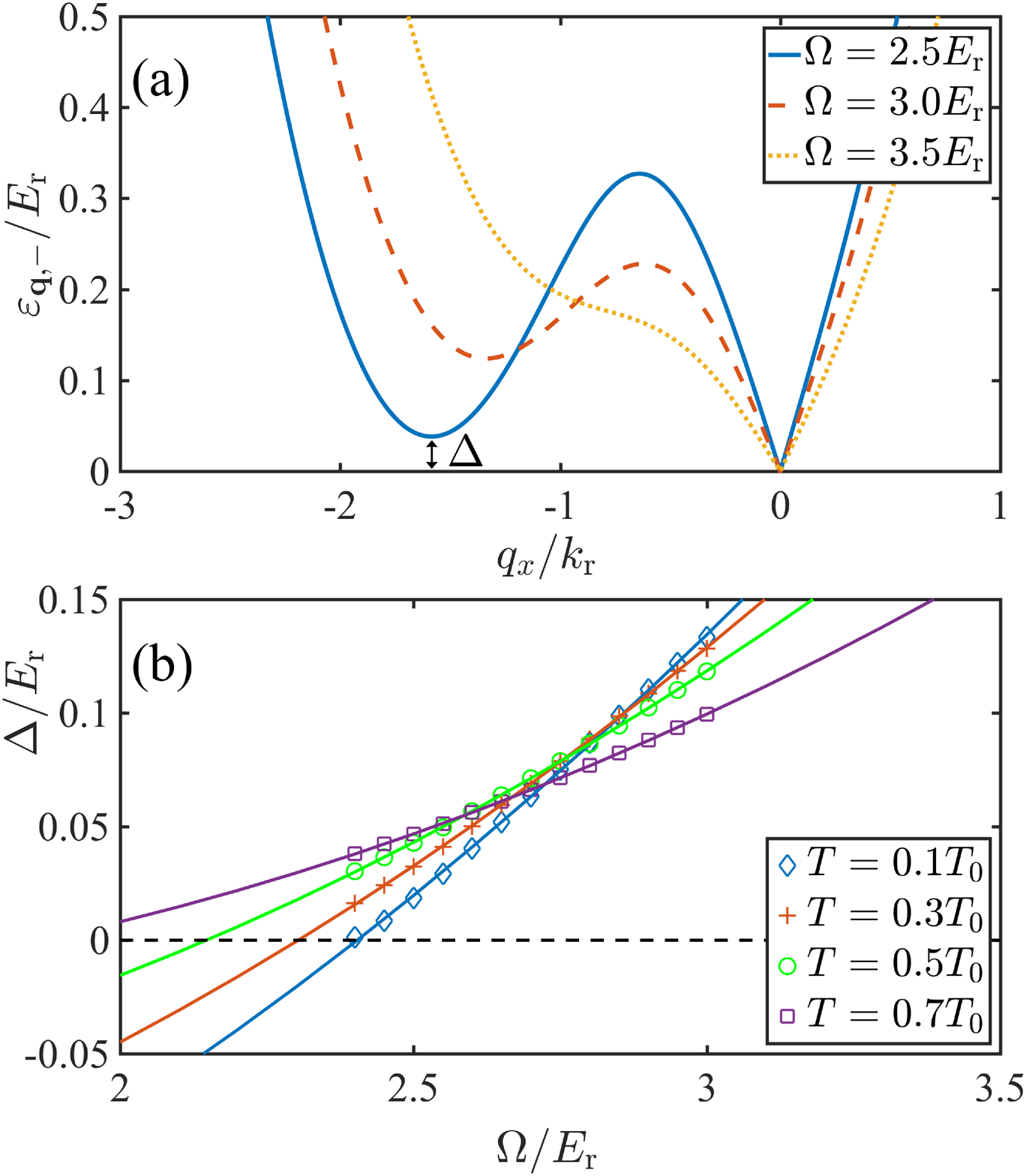} \caption{\label{fig2_soc} (color online). (a) The Bogoliubov excitation spectra
$\varepsilon_{\mathbf{q},\tau=-}$ at temperature $T=0.4T_{0}$ and
at various Rabi frequencies. Here, we set $q_{_{y}}=q_{z}=0$. The
roton gap at $\Omega=3.5E_{\textrm{r}}$ is explicitly indicated.
(b) The roton gap $\Delta$ as a function of Rabi frequency $\Omega$
at different temperatures, fitted with second-order polynomials (solid
lines).}
\end{figure}

Once the generalized GP and Bogoliubov equations are self-consistently
solved at finite temperature for a given set of variational parameters
($\theta,P_{x}$) \cite{note1}, we calculate straightforwardly the
free energy of the system, $\mathcal{F}(\theta,P_{x})/V=\mu n+\Omega_{G}/V$,
where the grand potential per volume is given by \cite{PitaevskiiStringariSBook,FetterWaleckaBook},
\begin{equation}
\frac{\Omega_{G}}{V}=(E_{0}-\mu)n_{c}+\frac{k_{B}T}{V}\sum_{{\bf q},\tau=\pm}\ln(1-e^{-\beta\varepsilon_{{\bf q}\tau}})
\end{equation}
and the condensate energy per particle is $E_{0}(\theta,P_{x})=(P_{x}^{2}+k_{\textrm{r}}^{2}-2P_{x}k_{\textrm{r}}\cos{2\theta})/(2m)+(gn_{c}-\Omega\sin{2\theta})/2-(g-g_{_{\uparrow\downarrow}})n_{c}\sin^{2}\theta\cos^{2}\theta$
\cite{Li2012PRL,Martone2012}. The two variational parameters $\theta$
and $P_{x}$ are determined by minimizing the free energy $\mathcal{F}$:
\begin{equation}
\left(\frac{\partial\mathcal{F}}{\partial\theta}\right)_{N}=0,~~\left(\frac{\partial\mathcal{F}}{\partial P_{x}}\right)_{N}=0.
\end{equation}
At zero temperature, under the approximation of negligible quantum
depletion (i.e., $n_{c}=n$), we have $\mathcal{F}=E_{0}(\theta,P_{x})nV$,
so the minimization can be carried out analytically \cite{Li2012PRL,Martone2012,Zheng2013}.
For $^{87}$Rb atoms with the interaction energies lengths $gn=0.38E_{\textrm{r}}$
and $g_{_{\uparrow\downarrow}}/g=100.99/101.20$ at typical density
$n=0.46k_{\textrm{r}}^{3}$ \cite{Ji2015}, this minimization leads
to two critical Rabi frequencies, $\Omega_{c1}\simeq0.2E_{\textrm{r}}$
and $\Omega_{c2}\simeq4.0E_{\textrm{r}}$, which locate the first-order
ST-PW and second-order PW-ZM transitions at zero temperature, respectively
\cite{Lin2011,Ji2014,Ji2015}. The phase space for the stripe phase
is therefore very narrow \cite{Ji2014}. To amplify the interaction
effects on the phase diagram, in all our numerical calculations, we
use a total density $n=1.0k_{\textrm{r}}^{3}$, the interaction energies
$gn=0.8E_{\textrm{r}}$ and $g_{_{\uparrow\downarrow}}n=0.5E_{\textrm{r}}$.
The relatively large difference in the intra- and inter-species interaction
strengths gives rise to a more experimentally accessible critical
Rabi frequency $\Omega_{c1}\simeq2.4E_{\textrm{r}}$ at $T=0$. We
note that, in the latest experiment on SOC Bose gases, where the two
spin-components are realized by two low-lying bands in a superlattice,
the interaction energy $g_{_{\uparrow\downarrow}}n$ can be tuned
at will by controlling the overlap in wave-functions of the two bands,
leading to a large phase space for the stripe phase with $\Omega_{c1}\simeq1.5E_{\textrm{r}}$
\cite{Li2017}.

\begin{figure}[t]
\centering{}\includegraphics[width=0.48\textwidth]{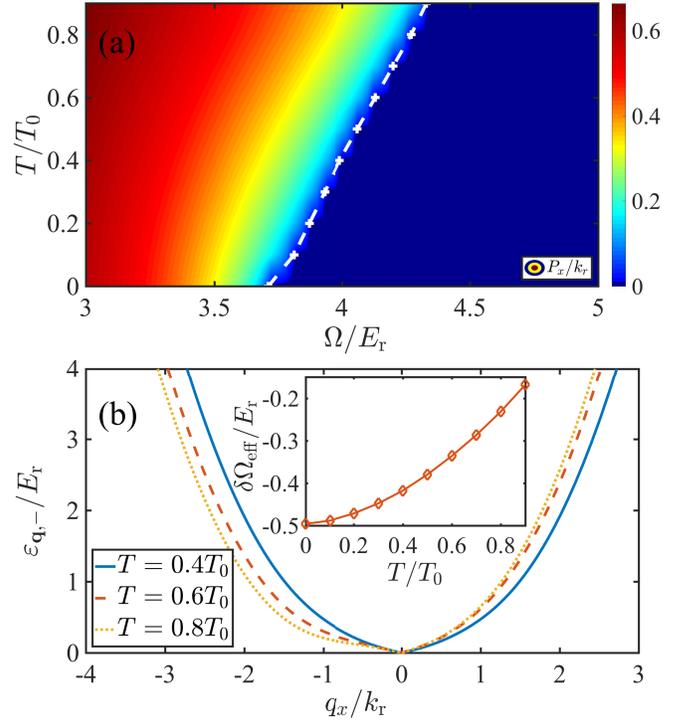}\caption{\label{fig3_soc} (color online). (a) The contour plot of the condensate-momentum
$P_{x}$, in units of $k_{\textrm{r}}$, as functions of $T$ and
$\Omega$. The boundary $P_{x}=0$ between PW and ZM phases is highlighted
by the white dashed line. (b) The Bogoliubov excitation spectra $\varepsilon_{\mathbf{q},\tau=-}$
at various temperatures, with $q_{_{x}}=q_{y}=0$. The inset shows
the temperature dependence of the effective Rabi frequency $\delta\Omega_{\textrm{eff}}=2g_{\uparrow\downarrow}\phi_{\uparrow}\phi_{\downarrow}$.
Here, we take $\Omega=4.0E_{\textrm{r}}$.}
\end{figure}

\textit{Phase diagram}. We are now ready to address the finite temperature
effects on the ST-PW and PW-ZM phase transitions. An intriguing feature
of the PW phase is the emergence of the roton-maxon structure in the
lowest-lying Bogoliubov excitation spectrum \cite{Zheng2012,Martone2012,Zheng2013},
as reported in Fig. \ref{fig2_soc}(a) at a nonzero temperature $T=0.4T_{0}$,
where $T_{0}=2\pi\hbar^{2}[(n/2)/\zeta(3/2)]^{2/3}/(mk_{B})$ is the
critical temperature of an ideal Bose gas with density $n/2$ in the
absence of SOC. This structure becomes much more pronounced with decreasing
Rabi frequency $\Omega$ and can be quantitatively characterized by
a roton gap $\Delta$, as indicated in Fig. \ref{fig2_soc}(a). Toward
the ST-PW transition, the roton gap is gradually softened and approaches
zero precisely at the transition. Therefore, as shown in Fig. \ref{fig2_soc}(b),
we determine the critical Rabi frequency $\Omega_{c1}$ from the $\Omega$-dependence
of the roton gap. As the accuracy of our numerical calculations becomes
worse near the transition, we typically fit the data by using a second-order
polynomial \cite{note2} and then calculate $\Omega_{c1}$ from the
fitting curve. By repeating this procedure at different temperatures,
we obtain the temperature dependence of the ST-PW boundary, as shown
in Fig. \ref{fig1_soc} by the empty circles.

On the other hand, we can straightforwardly determine the PW-ZM transition
by identifying the critical Rabi frequency $\Omega_{c2}$, at which
the condensate momentum $P_{x}$ approaches zero. This is illustrated
in the contour plot Fig. \ref{fig3_soc}(a) on the $T$-$\Omega$
plane, where the transition is highlighted by the white dashed line
(see also the empty diamonds in Fig. \ref{fig1_soc}). Furthermore,
in Fig. \ref{fig3_soc}(b), we check the temperature dependence of
the lowest-lying Bogoliubov excitation spectrum at $\Omega=4.0E_{\textrm{r}}$.
By increasing temperature from $0.4T_{0}$ to $0.6T_{0}$, and then
to $0.8T_{0}$, the spectrum changes from a symmetric shape (with
respect to $q_{x}=0$) to an asymmetric one. An asymmetric phonon
dispersion near $q_{x}=0$ is another characteristic feature of the
PW phase. It leads to different sound velocities when a density fluctuation
propagates along or opposite to the positive $x$-axis, which we shall
discuss in detail later.

From the phase diagram Fig. \ref{fig1_soc}, it is evident that the
phase space of the PW phase is greatly enlarged by temperature or
thermal fluctuations. The stripe phase is not energetically favorable
at relatively large temperature, as anticipated. As mentioned earlier,
the stripe phase is driven by the difference in the intra- and inter-species
interaction energies, i.e., $gn_{c}$ and $g_{\uparrow\downarrow}n_{c}$.
This difference becomes increasingly smaller with increasing temperature,
since the condensate density decreases. Hence, the stripe phase shrinks.
Our HFBP finding is consistent with the previous experimental determination
of the ST-PW boundary with $^{87}$Rb atoms at nonzero temperature
\cite{Ji2014}, and it also provides a useful \emph{microscopic} confirmation
of the perturbative theory by Yu \cite{Yu2014}. In contrast, the
significant shrinkage of the ZM phase at finite temperature - observed
as well with parameters for $^{87}$Rb atoms - is entirely not expected
(see, for example, the naive phase diagram sketched in Ref. \cite{Ji2014}).
Recall that the PW-ZM transition is largely due to the change of the
single-particle dispersion with increasing Rabi frequency \cite{Lin2011,Li2012PRL}.
From the mean-field point of view, therefore, the notable shift of
the PW-ZM boundary suggests a strong temperature dependence of the
\emph{effective} Rabi frequency $\delta\Omega_{\textrm{eff}}\equiv2g_{_{\uparrow\downarrow}}\phi_{\uparrow}\phi_{\downarrow}=-g_{\uparrow\downarrow}n_{c}\sin2\theta$,
which is resulted from the inter-species interaction (see Eqs. (\ref{eq:bogo1}),
(\ref{eq:bogo2}) and (\ref{eq: VarAs})). Indeed, as shown in the
inset of Fig. \ref{fig3_soc}(b), such a sensitive temperature dependence
is confirmed at a typical Rabi frequency $\Omega=4.0E_{\textrm{r}}$.

At high temperature, the SOC Bose gas becomes normal. Unfortunately,
sufficiently close to the superfluid-normal phase transition, our
HFBP theory becomes less accurate \cite{Griffin1996}. To determine
the transition temperature, we follow the previous work by Zheng \textit{et
al.} and adopt the Hartree-Fock approximation \cite{Zheng2013}. The
resulting critical temperature $T_{c}$ is shown at the top of Fig.
\ref{fig1_soc} by a solid curve with asterisks. At small Rabi frequency,
the predicted $T_{c}$ differs slightly from the classical Monte Carlo
simulation \cite{Arnold2001,Kashurnikov2001} (red cross), which confirmed
a linear shift of $T_{c}$ in the \textit{s}-wave scattering length
$\Delta T_{c}/T_{0}\approx1.3n^{1/3}a$. At large Rabi frequency (i.e.,
$\Omega>4E_{\textrm{r}}$), it is interesting that the Hartree-Fock
transition temperature matches very well with the PW-ZM boundary.
This is a strong indication that our HFBP predictions might be reliable
at temperature $T<0.9T_{0}$.

\begin{figure}[t]
\centering{}\includegraphics[width=0.48\textwidth]{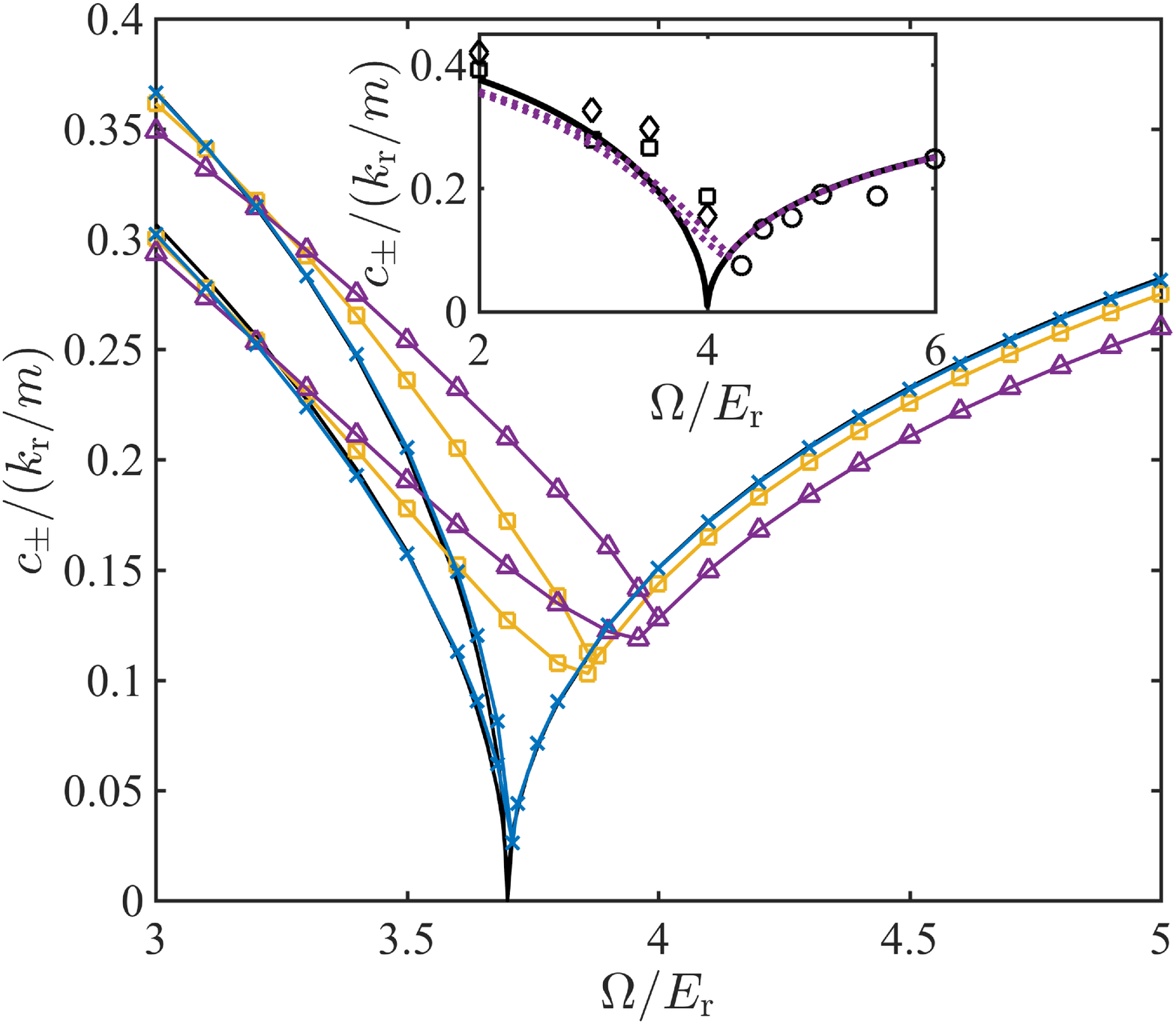}\caption{\label{fig4_soc} (color online) Sound velocities $c_{+}$ and $c_{-}$
as a function of Rabi frequency $\Omega$ at $T=0$ (black lines -
without quantum fluctuations; blue lines with crosses - with quantum
fluctuations), $T=0.2T_{0}$ (yellow lines with squares), and $T=0.4T_{0}$
(purple lines with triangles). The inset shows the sound velocities
in a $^{87}$Rb SOC gas at $T=0$ (black lines) and $T=0.3T_{0}$
(purple dotted lines). The symbols are the experimental data \cite{Ji2015}.
To simulate the experiment, we take a total density $n=0.46k_{\textrm{r}}^{3}$,
$gn=0.38E_{\textrm{r}}$ and $g_{_{\uparrow\downarrow}}/g=100.99/101.20$.}
\end{figure}

\textit{Sound velocities}. We now turn to discuss in greater detail
the anisotropic phonon dispersion at small $q_{x}$ in the PW phase
and the resulting two sound velocities $c_{+}>c_{-}$ \cite{Li2012PRL,Martone2012,Zheng2013},
which have been measured experimentally to characterize the PW-ZM
transition \cite{Ji2015}. Note that, turning into the ZM phase the
two velocities would merge into one, i.e., $c_{+}=c_{-}\equiv c$
\cite{Li2012PRL,Martone2012,Zheng2013}. 

In Fig. \ref{fig4_soc}, we report the behavior of sound velocities
across the PW-ZM transition as a function of Rabi frequency at various
temperatures. At zero temperature, all the previous studies predicted
a vanishing velocity right at the transition point $\Omega_{c2}$.
Physically, since the sound velocity may be well approximated by $c\simeq\sqrt{gn/m_{\mathrm{eff}}}$,
the vanishing sound velocity is a consequence of the flatness of the
single-particle spectrum at the transition and hence a divergent effective
mass $m_{\textrm{eff}}\rightarrow\infty$ \cite{Martone2012,Zheng2013}.
This interesting feature is exactly produced by our numerical calculation
if we do not account for the feedback of quantum fluctuations to the
total density (see the black solid curve). However, once we take into
account quantum fluctuations, there is a qualitatively change. Although
the sound velocities still exhibit a minimum at the transition, the
minimum becomes nonzero. This unexpected gap in sound velocity opened
by the enhanced quantum depletion at $\Omega_{c2}$ is typically about
$0.03k_{\textrm{r}}$ and might be detectable experimentally. A nonzero
temperature brings even more dramatic changes. As temperature increases
to $T=0.2T_{0}$ (the yellow curves with squares) and to $T=0.4T_{0}$
(the purple curves with triangles), we find that the minimum point
of sound velocity is progressively shifted toward larger Rabi frequency,
along with the shifted phase boundary $\Omega_{c2}$. At the same
time, the opening gap at the minimum is significantly enlarged by
thermal fluctuations.

To connect with the recent measurement for $^{87}$Rb atoms, we perform
a calculation by taking the peak density of the trapped cloud $n=n(\mathbf{r}=0)=0.46k_{\textrm{r}}^{3}$,
which leads to $gn=0.38E_{\textrm{r}}$ and $g_{\uparrow\downarrow}n=0.998\times0.38E_{\textrm{r}}$
\cite{Ji2015}. As shown in the inset of Fig. \ref{fig4_soc}, our
result at a realistic experimental temperature $T=0.3T_{0}$ (purple
dotted curves) agrees reasonably well with the measured sound velocities
(open symbols). The experimental data show a nonzero minimum or gap
at $\Omega_{c2}\simeq4.3E_{\textrm{r}}$ \cite{Ji2015}. The previous
theoretical studies at zero temperature instead predict a vanishing
sound velocity at $\Omega_{c2}\simeq4.0E_{\textrm{r}}$ (see the black
solid curves) and therefore fail to explain the observed shift of
the minimum, $\Delta\Omega_{c2}\sim0.3E_{\textrm{r}}$, and the gap
opening. It was suggested that the suppressed third spin state of
$^{87}$Rb atoms in the experiment may be responsible for the shift
\cite{Ji2015}. However, the finite gap remains to be explained. The
good agreement between our theory and experiment indicates that actually
the nonzero temperature in the experiment plays a crucial role near
the PW-ZM transition and it has to be accounted for in future experimental
investigations.

\textit{Conclusions}. In summary, a Hartree-Fock-Bogoliubov-Popov
theory has been developed to investigate the finite-temperature phase
diagram of a weakly-interacting spin-orbit coupled Bose gas in three
dimensions. We have shown that quantum and thermal fluctuations play
a significant role in enlarging the phase space for the plane-wave
phase. They also change the qualitative behavior of sound velocity
near the transition from the plane-wave phase to the zero-momentum
phase, by shifting the velocity minimum and inducing a finite gap.
For rubidium-87 atoms, our prediction on sound velocity at finite
temperature agrees qualitative well with the recent experimental measurement
and therefore provides a reasonable explanation for the puzzling observation
of gap opening \cite{Ji2015}. Further researches could be undertaken
to thoroughly explore the finite-temperature effects related to other
physical observables, such as anisotropic superfluid density and critical
velocity \cite{Yu2017}. Moreover, it is interesting to investigate
the exotic stripe phase, which may further reveal the intriguing supersolid
properties with cold atoms \cite{Li2017,Schirber2011}.
\begin{acknowledgments}
We thank very much Zeng-Qiang Yu, Si-Cong Ji and Long Zhang for many
discussions and for kindly providing their data in Refs. \cite{Zheng2013,Ji2015}.
XLC acknowledges the fruitful discussions with Jia Wang and the use
of supercomputer Green II at Swinburne \cite{SwinburneGreenII}. This
work was supported by the ARC Discovery Projects: DP140100637 and
FT140100003 (XJL), FT130100815 and DP140103231 (HH).
\end{acknowledgments}

\textit{Note added.} After this work was completed, we became aware
of two related works for a two-dimensional SOC Bose gas at finite
temperature: (i) A classical Monte Carlo simulation \cite{Kawasaki2017},
which addressed the Berenzinskii-Kosterlitz-Thouless (BKT) superfluid
phase transition induced by an anisotropic SOC, and (ii) a simulation
with the stochastic projected GP equation \cite{Su2017}, which revealed
a true long-range order hidden in the relative phase sector, coexisting
with the quasi-long-range BKT order in the total phase sector.


\begin{thebibliography}{10}
\bibitem{note1}We note that, here the self-consistency means that
the chemical potential $\mu$ or condensate density $n_{c}$ in the
coupled GP and Bogoliubov equations should be solved iteratively to
satisfy the number equation: $\sum_{\sigma}(n_{c\sigma}+n_{t\sigma})=n$,
where the total density $n$ is a given parameter at the beginning.
This iteration procedure, together with the minimization of the free
energy with respect to the variational parameter $\theta$ and $P_{x}$,
significantly increases the complexity of our numerical calculations.
\end{thebibliography}
\end{document}